# Spontaneous time reversal symmetry breaking in the pseudogap state of high-$T_c$ superconductors

A. Kaminski *†, S. Rosenkranz *†, H. M. Fretwell ‡, J. C. Campuzano *†, Z. Li §, H. Raffy §, W. G. Cullen †, H. You †, C. G. Olson ||, C. M. Varma ¶, and H. Höchst #

*Department of Physics, University of Illinois at Chicago, Chicago, Illinois 60607, USA

†Materials Science Division, Argonne National Laboratory, Argonne, Illinois 60439, USA

‡Department of Physics, University of Wales Swansea, Swansea, UK

§Laboratoire de Physique des Solides, Université Paris-Sud, 91405 Orsay, France

||Ames Laboratory, Iowa State University, Ames, Iowa 50011, USA

¶Bell Laboratories, Lucent Technologies, Murray Hill, New Jersey, 07974, USA

#Synchrotron Radiation Center, Stoughton, Wisconsin 53589, USA

**When matter undergoes a phase transition from one state to another, usually a change in symmetry is observed, as some of the symmetries exhibited are said to be spontaneously broken. The superconducting phase transition in the underdoped high-$T_c$ superconductors is rather unusual, in that it is not a mean-field transition as other superconducting transitions are. Instead, it is observed that a pseudo-gap in the electronic excitation spectrum appears at temperatures $T^*$ higher than $T_c$, while phase coherence, and superconductivity, are established at $T_c$ (Refs. 1, 2). One would then wish to understand if $T^*$ is just a crossover, controlled by fluctuations in order which will set in at the lower $T_c$ (Refs. 3, 4), or whether some symmetry is spontaneously broken at $T^*$ (Refs. 5-10). Here, using angle-resolved photoemission with circularly polarized light, we find that, in the pseudogap state, left-circularly polarized photons give a different photocurrent than right-circularly polarized**

**photons, and therefore the state below $T^*$ is rather unusual, in that it breaks time reversal symmetry[11]. This observation of a phase transition at $T^*$ provides the answer to a major mystery of the phase diagram of the cuprates. The appearance of the anomalies below $T^*$ must be related to the order parameter that sets in at this characteristic temperature .**

We have investigated the time reversal invariance of the electronic states by ARPES, which becomes sensitive to this symmetry by the use of circularly polarized photons[11]. The measured ARPES intensity $I_\alpha \propto |M_\alpha|^2$, where the matrix element $M_\alpha = \langle \mathbf{p}|O_\alpha|\psi(\mathbf{k})\rangle$, describes the ejection of the electron from an initial state $|\psi(\mathbf{k})\rangle$ to a final state $|\mathbf{p}\rangle$, and the dipole operator $O_\alpha$ contains the vector potential of $\alpha = L$ (left) or $\alpha = R$ (right) circularly polarized photons. The experimental setup is shown in Fig. 1. It consists of a plane grating monochromator beamline at the Aladdin synchrotron as a source of linearly polarized photons, quadruple reflection polarizer[12], refocusing mirror, and the experimental chamber. It is crucial in this experiment, which is essentially measuring absolute intensity changes, to minimize beam movement, as extraneous intensity changes can occur from such movements. We therefore monitor the beam position, as shown in Fig. 1, finding a small residual beam movement of 150 μm, which is compensated for by adjusting the experimental chamber position as the polarizer is rotated. In the experiment one wishes to maximize the product $TP^2$, where $T$ denotes the transmission and $P$ the polarization. In our experiment, this is accomplished with $P = 86\%$, as shown in Fig. 1d.

Thin film $Bi_2Sr_2CaCu_2O_{8+\delta}$ (Bi-2212) samples (1000Å –2000Å thick, *c*-axis perpendicular to the surface) of various dopings were grown using magnetron sputtering on a $SrTiO_3$ substrate. The choice of thin film samples was dictated by the requirement of absolute flatness. Another desirable property of these samples is the small superlattice signal, less than 3% in both ARPES and X-ray diffraction. Using synchrotron X-ray scattering we have verified that there are no symmetry changes (e.g. induced by the substrate) over the temperature range of the measurements as shown in Fig. 1c. The residual magnetization in the experimental chamber was 30 nT. The samples were cleaved in situ and measured at pressures $< 3 \times 10^{-11}$ Torr.

The symmetry properties of the ARPES matrix elements can be easily determined with respect to a mirror plane $m$ of the crystal, which is perpendicular to the surface. If one first considers, for simplicity, a time reversal symmetric initial state $|\psi(\mathbf{k})\rangle$, one finds two distinct cases: a) If the vector describing the propagation of the light $\hat{\mathbf{q}}_\gamma$, the normal to the surface $\hat{\mathbf{n}}$, and the final state momentum $\mathbf{p}$ are not all in $m$, then there is circular dichroism arising simply from the experimental geometry, due to the non-coplanarity of the three vectors in question. b) If the all three vectors $\hat{\mathbf{q}}_\gamma$, $\hat{\mathbf{n}}$, and $\mathbf{p}$ lie in $m$, then there is no dichroism for time-reversal invariant initial states.

Although case (b) is the one that interests us, we first describe case (a), as this is a large effect, and needs to be correctly accounted for in the analysis of the data[13-15]. In Fig. 2a we show the geometrical arrangement for case (a) when $\hat{\mathbf{q}}_\gamma$ and $\hat{\mathbf{n}}$ lie in the mirror plane $m$ diagonal to the $Cu$-$O$ bond direction, but the final state momenta $\mathbf{p}_1$ at point $M_1$, corresponding to $(\pi,0)$, and $\mathbf{p}_2$ at $M_2$ corresponding to $(0,\pi)$ are not in this plane. One can see in panels b and c of Fig. 2 that the intensity of the spectra at $M_1$ for the left-circularly polarized light is bigger than the one for the right-circularly polarized light. This situation is exactly opposite at the point $M_2$, as expected, since this point is a mirror reflection of the point $M_1$ about $m$ (Ref. 14). On the other hand, when the three vectors $\hat{\mathbf{q}}_\gamma$, $\hat{\mathbf{n}}$, and $\mathbf{p}$ are all in a mirror plane $m$, as illustrated in Fig. 2d where $m$ is the mirror plane along the $Cu$-$O$ bond direction, there is no dichroism (although the $Cu$-$O$ bond direction is strictly speaking not a symmetry direction of the crystal in the presence of a superlattice on the *Bi-O* plane, it was found experimentally[16], and confirmed here, that it *is* a symmetry direction for the *CuO$_2$* planar electronic states). This is seen in Fig. 2e, where we plot spectra obtained at point M$_1$ that is now in *m*. In this case the spectra have equal intensities to an accuracy of ± 0.06%, which sets the overall accuracy of the experiment. As a test, in Fig. 2f we also show spectra from polycrystalline Au, which also do not exhibit dichroism as the orientation of mirror planes is random.

The interesting situation arises when the initial state $|\psi(\mathbf{k})\rangle$ breaks time-reversal symmetry characterized by an order parameter θ. Therefore for **k** in some mirror planes *m*, $|\psi(\mathbf{k})\rangle$ is not an eigenstate of the reflection operator, even though the charge density retains the same lattice symmetries as for θ = 0. In this case, it is shown quite generally in Ref. 15 and for a particular case in Ref. 11 that there is a difference in photoelectron intensity for left (LCP) and right circularly polarized (RCP) light even for **k** in *m* (where the geometrical effect is absent). This effect, proportional to θ, is even for small variations of **k** about the mirror plane (while the geometric effect is odd). In order to experimentally detect such a TRSB state, one can plot the energy- integrated intensity[17] (-600 to 100 meV about the chemical potential) of the ARPES signal as a function of the momentum over a very small range (1/20 of the Brillouin zone) along a cut perpendicular to the mirror plane. In the absence of TRSB, the energy-integrated intensity as a function of momenta will be a straight line with the slope changing sign for the opposite circular polarization due to the odd symmetry of the geometrical effect. The two lines will cross at the mirror plane where both LCP and RCP intensities are equal. If upon cooling a TRSB state emerges, it will lead to a difference between the RCP and LCP at the symmetry plane.

In Fig. 3a we show the experimental geometry, while Fig. 3b shows the energy-integrated intensity $I_L$ and $I_R$ as a function of momentum along a cut indicated in panel a for an overdoped ($T_c$ = 64K) sample which *does not* exhibit a pseudogap[1,2], as seen from the energy spectra at $\mathbf{k}_f$ in panel d. In Fig. 3c we plot the dichroism signal $D = (I_R - I_L)/(I_R + I_L)$ obtained from the data shown in 3b. We see that the dichroism signal is independent of temperature and vanishes at the mirror plane, indicating the absence of TRSB in this sample (The non-zero $D$ away from the mirror plane is just the geometric effect discussed above).

We now perform the same measurements on an underdoped sample with a $T_c$ of 85K, which has a pseudogap and therefore *does* exhibit a change of the leading edge position of the photoemission spectrum with temperature as shown in Fig. 3h. The momentum dependence of the integrated intensity, Fig. 3f, shows that at high temperatures the LCP and RCP data cross at the mirror plane and therefore there is no TRSB. However, when the sample is cooled to below $T^*$ where the pseudogap opens, we observe a shift of the crossing point of the RCP and LCP data. This shift (2.3°) is much bigger than the temperature induced structural changes (0.05°), as seen in the rocking curves in Fig 1, and angular errors in the ARPES experiment (0.06°). We therefore have to conclude that now the intensities for LCP and RCP are *not* the same at the mirror plane by a difference much bigger than any experimental uncertainty. This is seen more clearly in Fig. 3g, which shows the relative difference $D$ for several temperatures. Each data point in Fig. 3 is the result of several measurements with alternating polarization directions. We further checked that subsequent warming of the sample through $T^*$ gave reproducible results as shown in Fig. 3g, which rules out sample aging as the origin of the effect. The constant value of the difference signal with energy (Fig. 3e) indicates that this effect is related to changes of the matrix elements and not due to artifacts arising from changes in the spectral function[17]. This observation of dichroism at the mirror plane provides direct evidence for TRSB in the pseudogap state of Bi-2212.

Additional information about the observed effect can be obtained by examining its symmetry. At $(\pi,0)$, the changes of $D$ are independent of **k** along the small cut perpendicular to the mirror plane (see Fig. 3g) and therefore the observed TRSB effect is *even* about the mirror plane along the *Cu-O* bond direction. Furthermore, from the dichroism data obtained from an underdoped sample ($T_c = 78K$) at two adjacent M points (i.e. $M_1 = (\pi,0)$ and $M_2 = (0,\pi)$) shown in Figs. 4a and 4b, we conclude that the effect is *odd* with respect to the mirror plane diagonal to the *Cu-O* bond direction. These data were obtained by first measuring the spectra at $M_1$ for T = 200K, then cooling the sample and measuring at T = 100K (Fig. 4a). While keeping the temperature constant, the sample was then rotated by 90° and $M_2$ was measured at low temperature. After this, the sample was warmed to 200K and measured again (Fig. 4b). At low temperatures, the RCP signal at $M_1$ is larger than the LCP signal and therefore $D$ is positive. At $M_2$ the opposite occurs, and $D$ is negative.

We have performed nine measurement sequences on four different underdoped and two overdoped samples. The sign and absolute value of the dichroism effect is different for different samples, which we attribute to a variation of the number of the two possible domains formed below $T^*$. Most significantly however, the effect is only observed in underdoped samples below $T^*$, never above. Our measurements on samples of different doping levels clearly establish that the effect is tied to the pseudogap line, as shown in Figs. 4c and 4d, leading us to conclude that time reversal symmetry is broken in the pseudogap state of Bi-2212. We also find that, for underdoped samples, the TRS breaking persists into the superconducting state.

Acknowledgements. We thank Mohit Randeria for many detailed discussions and T. Kubala, G. Rogers and M. Bissen of the Synchrotron Radiation Center for their help with the polarizer. This work was supported by the NSF and the US Dept of Energy - Basic Energy Sciences. The Synchrotron Radiation Center is supported by the NSF. SR is supported by the Swiss NSF.

**Correspondence and requests for materials should be addressed to J.C.C. (email: jcc@uic.edu).**

Figure 1. Schematic layout and accuracy of the experimental arrangement. **a**, Experimental setup. The energy resolution was chosen to be 30 meV. The anlyzer was operating in angle resolved mode, acquiring 21 Energy Distribution Curves (EDCs) over a range of 5°. The sample positioning precision was 25 µm. The samples were mounted in normal incidence geometry with a precision of 0.1° using the beamline integrated laser. During the experiment the angular position of the samples was maintained to within 0.06° using an external laser beam. **b**, Motion of the uv beam during rotation of the polarizer. **c**, X-ray diffraction of the Bi-2212 thin films, showing that there is no distortion of the $CuO_2$ planes and no rotation of the mirror plane along the *Cu-O* bonds to within 0.05°. **d**, Degree of circular polarization for RCP (circles) and LCP (triangles) as used in the experiments. For purely circular polarized light, E($\phi$) would be constant (dashed line), whereas for linear polarized light, E($\phi$) would take the form of an eight, being zero perpendicular to the axis of polarization.

Figure 2. Illustration of the geometric Dichroism in ARPES depending on the experimental setup. **a**, Geometric arrangement of the incident photon, final momentum of the photoemitted electron and symmetry axis when they are not all in the same plane. **b**, Spectra at the $M_1$ point with LCP and RCP polarizations from an optimally doped sample at T = 300K. **c**, Spectra at the $M_2$ point with LCP and RCP polarizations from the same sample as in **b**. **d**, Geometrical arrangement when the incident photon, final state momentum and symmetry direction are all in the same plane. **e**, Spectra at the M1 point with LCP and RCP,

showing the accuracy of the experiment of ±0.06% from an optimally doped sample at T = 300K. **f**, Spectra from polycrystalline Au with LCP and RCP.

Figure 3. Results of dichroism experiments in overdoped and underdoped samples. **a**, Geometrical arrangement for the experiment, where the incident photon direction, the final state momentum and the normal to the surface are all in the mirror plane. The left side shows an enlarged region around the (π,0) point and the range (red line) covered by the Detector. **b**, Energy integrated intensity as a function of momentum perpendicular to the mirror plane at the $M_1$ point of the BZ for LCP and RCP for an overdoped ($T_c$ = 64K) sample for T = 250K (red) and T = 100K (blue). The lines are linear fits to the data. The data were normalized to the second order. The error bars are statistical only and represent one standard deviation. **c**, Relative difference $D = (I_L-I_R) / (I_L+I_R)$ of the Data shown in **b** and in addition for T=200K (purple) and 150K (green), indicating no circular dichroism for a sample with no shift in the leading edge with temperature **d**. **e**, Difference signal as a function of temperature and energy (integrated over 50meV). **f**, Energy integrated intensity as a function of momentum perpendicular to the mirror plane at the $M_1$ point of the BZ for LCP and RCP for an underdoped sample ($T_c$ = 85K) that displays a shift in the leading edge (panel **h**), showing no dichroism for T = 250K but clearly showing that there is dichroim for T = 100K. **g**, Relative difference $D$ of the data shown in **f** and in addition for T = 200K and T = 150K. The diamond data points for T = 250K were taken at the end of the experiment after a cooling cycle.

Figure 4. Symmetry and temperature dependence of the observed dichroism. **a**, Relative difference measured for an underdoped sample at $M_1$ and **b**, after

rotating the sample by 90° at $M_2$. **c**, Temperature dependence of the dichroism signal at $M_1$ from the data shown in Fig. 3c (overdoped 64K) and 3f (underdoped 85K). **d**, Phase diagram of Bi-2212 showing the existence (red diamonds) and absence (blue dots) of dichroism at the mirror plane along the *Cu-O* bonds.

figure 1

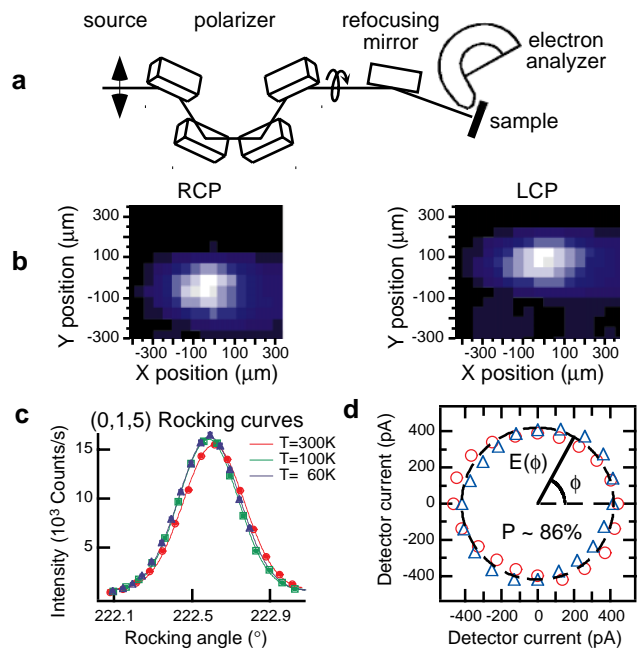

figure 2

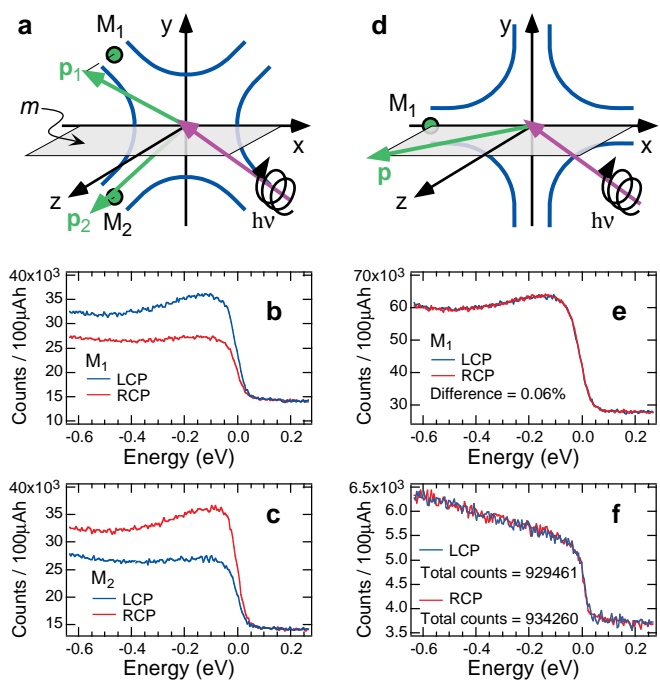

figure 3

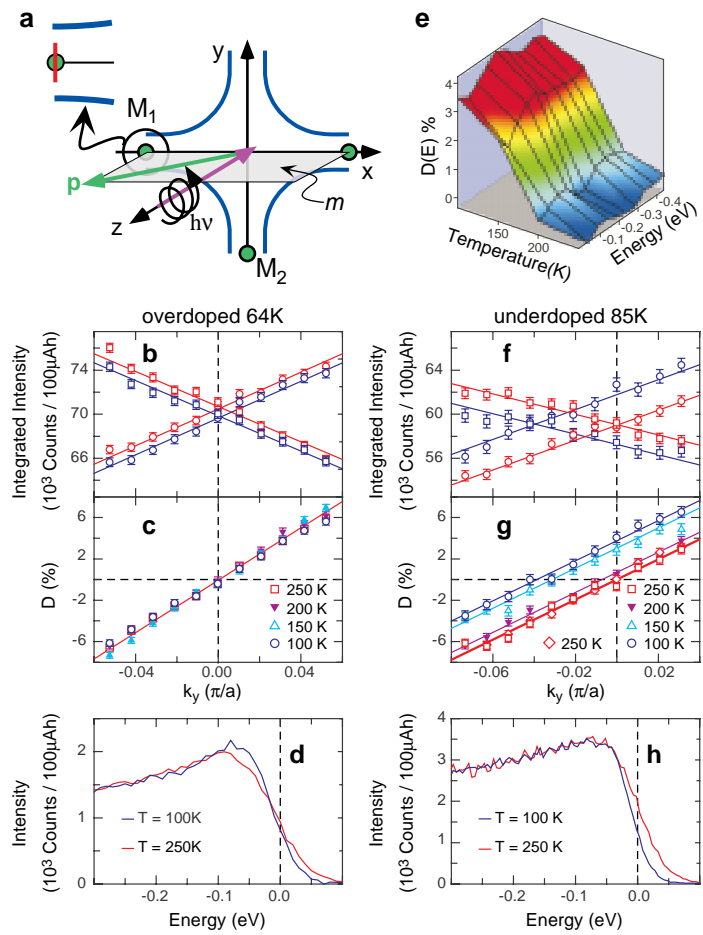

figure 4

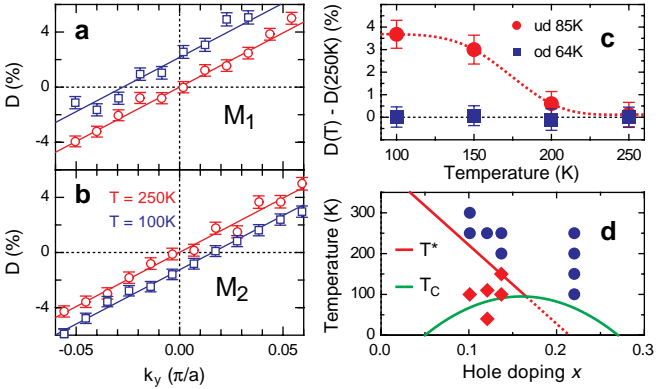